\documentclass[useAMS,usenatbib]{mn2e}
\onecolumn
\usepackage{lineno}
\usepackage[dvipdfmx]{graphicx,color,hyperref}
 
\usepackage{hyperref}
\usepackage{threeparttable}
\usepackage{multirow}
\usepackage{ulem}
\usepackage{amsmath}
\usepackage{float}
\usepackage{graphicx}
\usepackage{txfonts}
\usepackage{indentfirst}
\usepackage{color}
\usepackage{ulem}
\usepackage[dvipdfmx]{hyperref}
\hypersetup{
 setpagesize=false,
 bookmarksnumbered=true,%
 bookmarksopen=true,%
 colorlinks=true,%
 linkcolor=green,
 citecolor=green,
}

\newcommand{\subaru}{{\it Subaru}}
\newcommand{\suzaku}{{\it Suzaku}}

\newcommand{\swift}{{\it Swift}}
\newcommand{\fermi}{{\it Fermi}}

\newcommand{\pasj}{PASJ}
\newcommand{\nat}{Nature}
\newcommand{\apj}{ApJ}
\newcommand{\apjl}{ApJL}

\newcommand{\aap}{A\&A}
\newcommand{\ssr}{Space Sci. Rev.}
\newcommand{\mnras}{MNRAS}

\newcommand{\grbA}{050904}
\newcommand{\grbB}{120521C}
\newcommand{\grbC}{130606A}
\newcommand{\TgrbB}{$T_{0, {\rm 120521C}}$}
\newcommand{\TgrbC}{$T_{0, {\rm 130606A}}$}

\newcommand{\z}{{\it z}}

\newcommand{\eiso}{$E_{\rm iso}$}
\newcommand{\epeak}{$E_{\rm peak}^{\rm obs}$}
\newcommand{\epeakrest}{$E_{\rm peak}^{\rm src}$}
\newcommand{\ek}{$E_{\rm K}$}
\newcommand{\egamma}{$E_{\gamma}$}
\newcommand{\lpeak}{$L_{\rm p}$}
\newcommand{\tjet}{$t_{\rm jet}$}
\newcommand{\thetajet}{$\theta_{\rm jet}$}

\newcommand{\erg}{erg}
\newcommand{\ergs}{${\rm erg\,s^{-1}}$}
\newcommand{\ergscm}{${\rm erg\,s^{-1}\,cm^{-2}}$}
\newcommand{\three}{I\hspace{-.1em}I\hspace{-.1em}I} % Third
\title[High-redshift GRBs, 120521C and 130606A]
{Hard X-ray Spectral Investigations of Gamma-ray Bursts, 120521C and 130606A, at High-redshift $z\sim6$}
\author[T. Yasuda, Y. Urata, J. Enomoto, and M. S. Tashiro]
{T. Yasuda$^{1}$\thanks{E-mail: yasuda@heal.phy.saitama-u.ac.jp},
Y. Urata$^{2}$, J. Enomoto$^{1}$, and M. S. Tashiro$^{1}$\\
$^{1}$Graduate School of Science and Engineering, Saitama University, 255 Shimo-Okubo, Sakawa, Saitama, Saitama, 338-8570, Japan\\
$^{2}$Institute of Astronomy, National Central University, Chung- Li 32054, Taiwan
}

\begin{document}
\date{Accepted ?????. Received ?????; in original form ?????}
\pagerange{\pageref{firstpage}--\pageref{lastpage}} \pubyear{2015}
\maketitle

\label{firstpage}

\begin{abstract}
This study presents the temporal and spectral analysis of the prompt emission of two high-redshift gamma-ray bursts (GRBs), 120521C at $z\sim6$ and 130606A at $z\sim5.91$, which were performed using the \swift-XRT/BAT and the \suzaku-WAM simultaneously.
Based on follow-up XRT observations, the longest durations of the prompt emissions were approximately $80$\,s (120521C) and $360$\,s (130606A) in the rest frame of each GRB, which are categorized as long-duration GRBs, but are insufficiently long compared with the predicted duration of GRBs that originate from first-generation stars.
Because of the wide bandpass of the instruments covering the ranges of 15\,keV--5\,MeV (BAT-WAM) and 0.3\,keV--5.0\,MeV (XRT-BAT-WAM), we successfully determined the $\nu F_{\nu}$ peak energies \epeakrest\ of $682^{+845}_{-207}$\,keV and $1209^{+553}_{-304}$\,keV in the rest frame, and the isotropic-equivalent radiated energies \eiso\ of $(8.25^{+2.24}_{-1.96})\times10^{52}$\,\erg\ and $(2.82^{+0.17}_{-0.71})\times10^{53}$\,\erg, respectively.
These obtained characteristic parameters are in accordance with the well-known relation between \epeakrest\ and \eiso\ (Amati relation).
In addition, we examined the relations between \epeakrest\ and the 1-s peak luminosity, \lpeak, or the geometrical corrected radiated energy, \egamma, and confirmed the \epeakrest-\lpeak\ (Yonetoku) and \epeakrest-\egamma\ (Ghirlanda) relations.
The results implied that these high-redshift GRBs at $z\sim6$, expected as having radiated from the reionization epoch, have similar properties as that of X-ray prompt emission with low-redshift GRBs.
\end{abstract}

\begin{keywords}
methods: data analysis --
gamma-ray burst: individual: \grbB --
gamma-ray burst: individual: \grbC
\end{keywords}

%%%%%%%%%%%%%%%%%%%
%%%%    INTRODUCTION    %%%%
%%%%%%%%%%%%%%%%%%%

\section{Introduction} \label{s-int}

Gamma-ray bursts (GRBs) at a high redshift are considered a powerful probe for the early universe at \z$\sim$6--20, and a portion of their progenitors are first-generation stars, called Population \three\ (Pop. \three) stars.
Radiations emitted from these stars have been predicted to cause a reionization state of the intergalactic medium (IGM).
Theoretical investigations of production of GRBs from Pop III stars have been intensively performed \citep[e.g.][]{Meszaros2010,Komissarov2010,suwa2011,Nagakura2012,woosley2012,nakauchi2012,toma2016}.
Most of models prefer to use the very massive progenitor stars with $M > 100$ solar mass, and predict the peculiarly large total energies and ultra-long duration \citep[e.g.][]{suwa2011}.
On the other hand, numerical studies of the evolution of Pop. \three\ stars imply that luminous GRBs are rather difficult to occur from the very massive progenitor stars \citep{Yoon2012, Yoon2015}.
For GRB originated from the Pop. \three\ stars with mass smaller than 100 solar mass (typically 40 solar mass), \citet{nakauchi2012} predicted the duration of $\sim10^{5}$ s in the observer frame.
Ultra-long duration with the time scale of $10^{3}$--$10^{7}$\,s is therefore the common expected property from GRBs originated from Pop III stars.
Although prompt spectral properties of Pop. \three\ stars were expected \citep[e.g.][]{nakauchi2012}, it fully relies on the correlations \citep{amati2002,yonetoku2004}, which are established using GRBs at lower redshift.
Hence, prompt emission characterizations of long GRBs at higher redshift could enrich the studies of Pop. \three\ stars. 

To date, a few GRBs, namely 050904 \citep[$z=6.295$,][]{Tagliaferri2005,haislip2006,kawai2006}, 080913 \citep[$z=6.70$,][]{greiner2009}, 090423 \citep[$z=8.1$,][]{salvaterra2009,tanvir2009}, 090429B \citep[$z\sim9.4$,][]{cucchiara2011}, 120521C \citep[$z\sim6$,][]{tanvir2012,laskar2014},130606A \citep[$z=5.91$,][]{chornock2013,castro-tirado2013,totani2014}, and 140515A \citep[$z=6.32$,][]{chornock2014,melandri2015}, have been identified as high-redshift GRBs at \z~$\gtrsim$ 6.
These brief properties are listed in Table~\ref{tab:0}.
In particular, according to \citet{zhang2014} duration of 050904 in X-ray band exceeds $3.2 \times 10^5$\,s, but the authors also suggested ultra-long GRBs were not outliers from the distribution of the duration of ordinary GRBs. 

According to the optical afterglow spectrum of \grbC\ obtained with the \subaru\ telescope \citep{totani2014}, the measured neutral fraction of the IGM implies that reionization is incomplete at $z\sim6$, which provides critical information on the reionization history of the cosmos.
This result is consistent with the neutral fraction of the IGM measured using quasar spectra at $z\sim6$ \citep{schroeder2013}, which is less uncertain.
We consider this implication to support the possibility that the Pop. \three\ star is a progenitor of a high-redshift GRB, which is expected to form with completely metal-free gases, the composition of which might be relatively different from present stars.
However, the correlations between the $\nu F_{\nu}$ peak energy in the X-ray spectrum and the radiated energy of prompt emissions are known, and GRBs are expected to be a useful tool for determining the cosmological parameters as a standard candle like type Ia supernovae.
GRBs have the strength of reaching a higher redshift compared with supernovae.
This raises the the following question; Could high-redshift GRBs produced from Pop. \three\ stars also fit the correlations confirmed with low-redshift GRBs at \z\ $\lesssim$ 6?

Fireball dissipation model in photosphere \citep{ioka2010} was implied by the \epeakrest-\lpeak\ (Yonetoku) relation as a candidate of radiation mechanism of the prompt emission.
Blackbody law in the surface of the photosphere follows $L \propto (r_{\rm ph}/\Gamma_{\rm ph})^{2}T^{4}$, where $r_{\rm ph}$ is radius of the photosphere; $\Gamma_{\rm ph}$ is a Lorentz factor of its radiation flow; and $T$ is blackbody temperature observed as $\nu F_{\nu}$ peak energy \epeakrest.
Considering Pop. \three\ stars are one-hundred times larger radius \citep{ohkubo2009,ioka2010} and luminosity $L$ and the Lorentz factor are comparable with those of Wolf-Rayet stars, Pop. \three\ stars have ten times lower \epeakrest\ than that of Wolf-Rayet stars is required.
We predicted that high-redshift GRBs deviate from Yonetoku relation as lower \epeakrest.
However, the two characteristic parameters of $\nu F_{\nu}$ peak energy \epeak\ and the isotropic-equivalent radiated energy \eiso\ have been measured in only one object (050904) to date, without any assumptions of X-ray spectral shapes (Table~\ref{tab:0}).
Thus, sensitive X-ray observations of the prompt interval of the GRBs with a broadband from the keV to MeV energy range are critical for measuring the two characteristic parameters.

This paper presents X-ray spectral analyses of the prompt emissions of two high-redshift GRBs, which were conducted using \suzaku\ wide-band all-sky monitor \citep[WAM,][]{yamaoka2009}, \swift\ burst alert telescope \citep[BAT,][]{barthelmy2005}, and \swift\ X-ray telescope \citep[XRT,][]{burrows2005}.
To better constrain the spectral peak energy \epeak\ and isotropic-equivalent energy \eiso, we performed simultaneous joint spectral fitting between BAT-WAM (nominal bandpass: 15\,keV--5\,MeV) and XRT-BAT-WAM (0.5\,keV--5\,MeV).
Because these combinations provide broadband spectral information, investigating these spectral features presents an advantage \citep[e.g.][]{krimm2009,urata2009,urata2014}.
Furthermore, the detailed spectral cross-calibration studies were performed among the BAT and the WAM \citep{sakamoto2011}.
The following sections detail our employment of FTOOLS provided in the HEADAS software package (version\ 6.14) for data reduction and analyses, and the quoted error was at a 90\% confidence level.
The cosmological parameters of $H_{0} = 70$\,km\,s$^{-1}$\,Mpc$^{-1}$, $\Omega_{m} = 0.3$, and $\Omega_{\Lambda} = 0.7$ were used.

%%%%%%%%%%%%%%%%
%%%%    Instruments    %%%%
%%%%%%%%%%%%%%%%

\section{Instruments and Analyses}\label{s-ins}

%----  WAM  ---%
The WAM is an active-shield of a hard X-ray detector \citep[HXD,][]{takahashi2007,kokubun2007} onboard \suzaku\ satellite \citep{mitsuda2007}.
It is composed of four Bismuth Germanate crystal walls (WAM-0, WAM-1, WAM-2, and WAM-3).
Although the main function of the WAM is exception of non-X-ray background events caused by cosmic-particles from observed data of main sensor of the HXD, the WAM is effective for monitoring transient astronomical events in the nominal bandpass range of 50\,keV--5\,MeV \citep{yamaoka2009}.
However, because of the long-term degradation of the gain during the observational period, the energy bandpass shifted to 88\,keV--7.7\,MeV on May 21, 2012, and 91\,keV--8.0\,MeV on June 6, 2013.
Because of its wide field of view of 2$\pi$\,st and large effective area of 400\,cm$^{2}$ for 1-MeV photons, the WAM can be used for investigating GRBs, solar flares, and magnetar bursts \citep[e.g.][]{endo2010,urata2012,yasuda2015,ohmori2016}.
The analytical tools applied to the WAM data, {\tt hxdmkwamlc} displaying a light curve and correcting dead time, and {\tt hxdmkwamspec} extracting energy spectrum.
Furthermore, we used {\tt battblocks} to calculate the $T_{90}$ duration, which is the time of accumulation between 5\% and 95\% of the counts.
In the following sections, we used these tools and a Monte-Carlo response generator \citep{ohno2005}, which was used to generate detector response matrices for the spectral analyses.

%---  BAT & XRT ---%
The BAT and the XRT onboard \swift\ satellite \citep{gehrels2004} are powerful detectors for the discovery and observation of transient astronomical objects.
Initial detections and localizations of \grbB\ and \grbC\ were also performed using the BAT \citep{baumgartner2012,ukwta2013}.
Both detectors provide detailed time-resolved temporal information compared with the WAM by using BAT trigger data and XRT windowed-timing (WT) mode data.
In addition, in-depth follow-up observations with the XRT provided the long-term light curves of GRBs.
BAT and XRT spectra cover bandpasses of 0.5--10\,keV and 15--150\,keV, respectively.

This paper presents the temporal and spectral analyses of a part of the prompt emission of the GRBs.
First, we used the reprocessing tools {\tt batgrbproduct} for the BAT trigger data sets and {\tt xrtpipeline} for the XRT WT data.
Second, we employed {\tt batbinevt} and {\tt batdrmgen} to extract the BAT energy spectra and generate these response matrices, respectively.
For the spectral analyses of the XRT, we used {\tt xselect} to accumulate XRT energy spectra.
We then extracted the source and background spectra within 20 pixels from the center of the location of the GRB and between 20--40 pixels, respectively.
We used a response matrix ``swxwt0to2s6\_20130101v015.rmf,'' which is available from the calibration database CALDB, and generated an ancillary response file by using {\tt xrtmkarf}.
In the following spectral analyses, we fixed the flux normalization of the BAT to 1, and maintained the factor of the WAM and XRT as a free parameter.

By using the three instruments simultaneously, we achieved four orders of broadband joint spectral fitting within an energy range of 0.5\,keV--5\,MeV, which is more advantageous for determining such spectral characteristics compared with the individual use of these instruments.

%%%%%%%%%%%%%%%%
%%%%    Observation    %%%%
%%%%%%%%%%%%%%%%

\section{Observations and Results}\label{s-obs}

To perform the spectral analyses on high-redshift GRBs, we listed seven high-redshift GRBs, as shown in Table~\ref{tab:0} (GCN circulars\footnote[1]{http://gcn.gsfc.nasa.gov} and GRB big data table\footnote[2]{http://www.mpe.mpg.de/~jcg/grbgen.html}).
All listed GRBs triggered the BAT and were localized by it.
Previous spectral studies on four GRBs (050904, 080913, 090423, and 140515A) have reported the \epeak\ and \eiso\ values \citep{sugita2009,yonetoku2010,vonKienlin2009,melandri2015}, although the two characteristics (except for 050904) were derived under the assumption of one or two model parameters of the Band function \citep{band1993} of $\alpha$ or $\beta$, or by using a power-law (PL) with an exponential cut-off (PLE) model.
Because of the broadband spectrum between the BAT and WAM, \citet{sugita2009} successfully determined the two characteristics for 050904 without the requiring the aforementioned assumption.

The 090429B recorded a photometric redshift of $z\sim9.4$ \citep{cucchiara2011}, which is the greatest distance observed among GRBs to date.
However, only \swift\ satellite observed its prompt emission, and the two spectral characteristic parameters have not been determined because of the limited bandwidth \citep{cucchiara2011}. 
Regardless, the remaining two GRBs, \grbB\ and \grbC\, were successfully simultaneously observed with the WAM and BAT.
We conducted the broadband spectral analyses as follows:

%%%  GRB 120521C  %%%
\subsection{GRB\grbB}\label{s-anaB}

%---  GRB 120521C  ---%
The BAT triggered \grbB\ \citep{baumgartner2012,markwardt2012} on May 21, 2012, at 23:22:07.703 (UT) ($=$\TgrbB, hereafter), and localized its position at (RA, Dec)(J2000) = (14h17m09s, + 42 d 07' 23")
Afterglow emissions of the GRB were successfully observed in multi-wavelength from radio, near infrared, optical to X-ray, and a jet-break time \tjet\ of $\sim7$\,d was revealed also in \citet{laskar2014}.
According to the paper, a jet opening angle \thetajet$=0.052^{+0.040}_{-0.019}$\,rad, a beaming-corrected kinetic energy \ek$=(3.1^{+1.9}_{-0.9}) \times 10^{50}$\,\erg, and a corrected radiated gamma-ray energy \egamma$=(2.6^{+4.4}_{-2.0})\times\,10^{50}$\,\erg\ were derived using Markov Chain Monte Carlo analysis.
Then its redshift was estimated \z~$\sim6$.
However, the energies (1\,keV--10\,MeV) were calculated from only observed BAT spectrum (15-150\,keV).
Therefore, more broader bandpass spectrum than that of only the BAT is important to reveal the energetics.

\subsubsection{Light Curves of \grbB}\label{s-anaB-l}

\begin{figure}
\includegraphics[width=84mm]{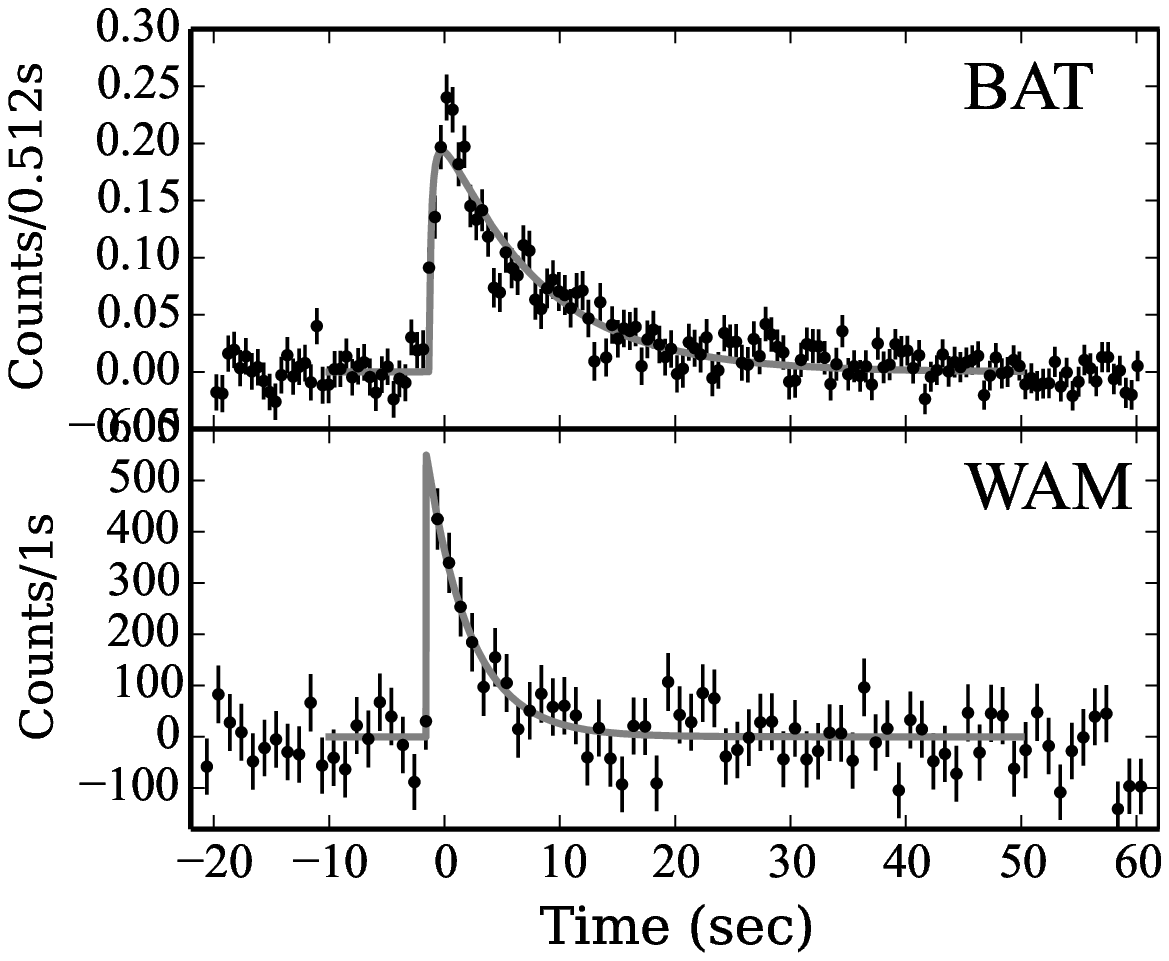}
\includegraphics[width=84mm]{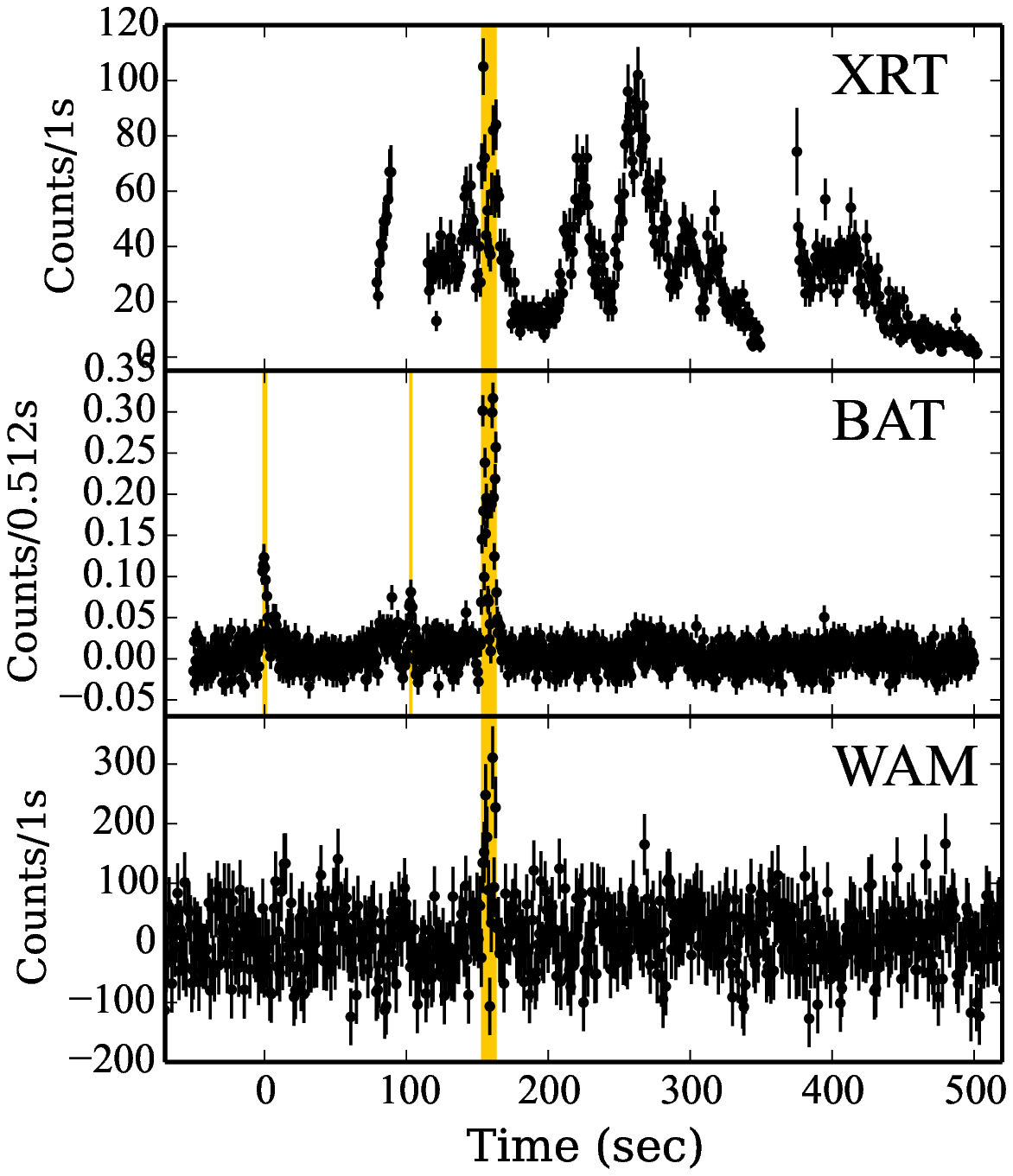}
\caption{
Light curves of the prompt emissions of \grbB\ (left) and \grbC\ (right), which are accumulated with 1\,s for WAM and XRT, and 512\,ms for BAT.
Background components of the WAM light curves were subtracted.
Gray lines represent FRED functions obtained in \S\ref{s-anaB}.
Orange hatched regions are extracted intervals as region~A, B, and C.
}
\label{fig:1}
\end{figure}

After the BAT detection of \grbB\ at \TgrbB, \swift\ satellite immediately slewed, and the XRT began conducting observations in WT mode.
The WAM also detected the GRB, which was recognized by a ground-burst search at 23:22:06.595 (UT) with the 1-s time-resolution data of WAM-1.
%---  Light Curves  ---%
As shown in the left-hand panel of Figure~\ref{fig:1}, the light curves of the prompt part of the GRB exhibited a fast rise and exponential decay (FRED)-like pulse.
The $T_{90}$ duration in the WAM light curve was measured as $10 \pm 4$\,s.
However the observed duration itself may not indicate of central engine activity directly, since it is strongly affected by sensitivity of the observation instruments as well as the redshift and time dilation effect.
We therefore estimated the duration of the GRB using X-ray light curves as following paragraphs.

The FRED function is indicated by \citet{norris2005,peng2010} as $f(t) = A \lambda\, \mathrm{exp}[-\tau_{1}/(t-t_{\rm s})-(t-t_{\rm s})/\tau_{2}]$, where $t$ is time since \TgrbB; $A$ represents the normalization of the pulse height; $t_{\rm s}$ is the start time; $\tau_{1}$ and $\tau_{2}$ denote the rise and decay characteristics, respectively; and $\lambda$ is defined as ${\rm exp}[2(\tau_{1}/\tau_{2})]^{1/2}$.
According to the previous researches, by using the two parameters of $\tau_{1}$ and $\tau_{2}$, we can calculate a pulse width between the two $1/e$ intensity points, $w = \Delta \tau_{1/e} = \tau_{2}(1+2{\rm ln}\lambda)^{1/2}$, and a pulse asymmetricity, $\kappa = \tau_{2}/w$.
Furthermore, the rise and decay constant are derived from the two characteristics as $\tau_{\rm rise} = 1/2 \cdot w (1-\kappa )$ and $\tau_{\rm dec} = 1/2 \cdot w (1+\kappa )$, respectively.
The two light curves (Figure~\ref{fig:1}) fit well with the FRED functions.
Table~\ref{tab:1} shows the obtained test statistics and the FRED characteristic parameters.
Because several previous studies on FRED GRBs \citep[e.g.][]{peng2012,tashiro2012,tashiro2014} have indicated the energy dependency of the decay constant to be $\tau_{\rm dec}(E) \propto E^{\delta_{\tau_{\rm dec}}}$, we examined the fit of the FRED function as a narrower energy bandpass, as shown in Table~\ref{tab:1}.
Thereafter, above the 409\,keV band of the WAM light curve, we could not observe a significant pulse, and thus, the light curve was divided into two bandpasses, 88--195\,keV and 195--409\,keV.

\begin{figure}
\includegraphics[width=84mm]{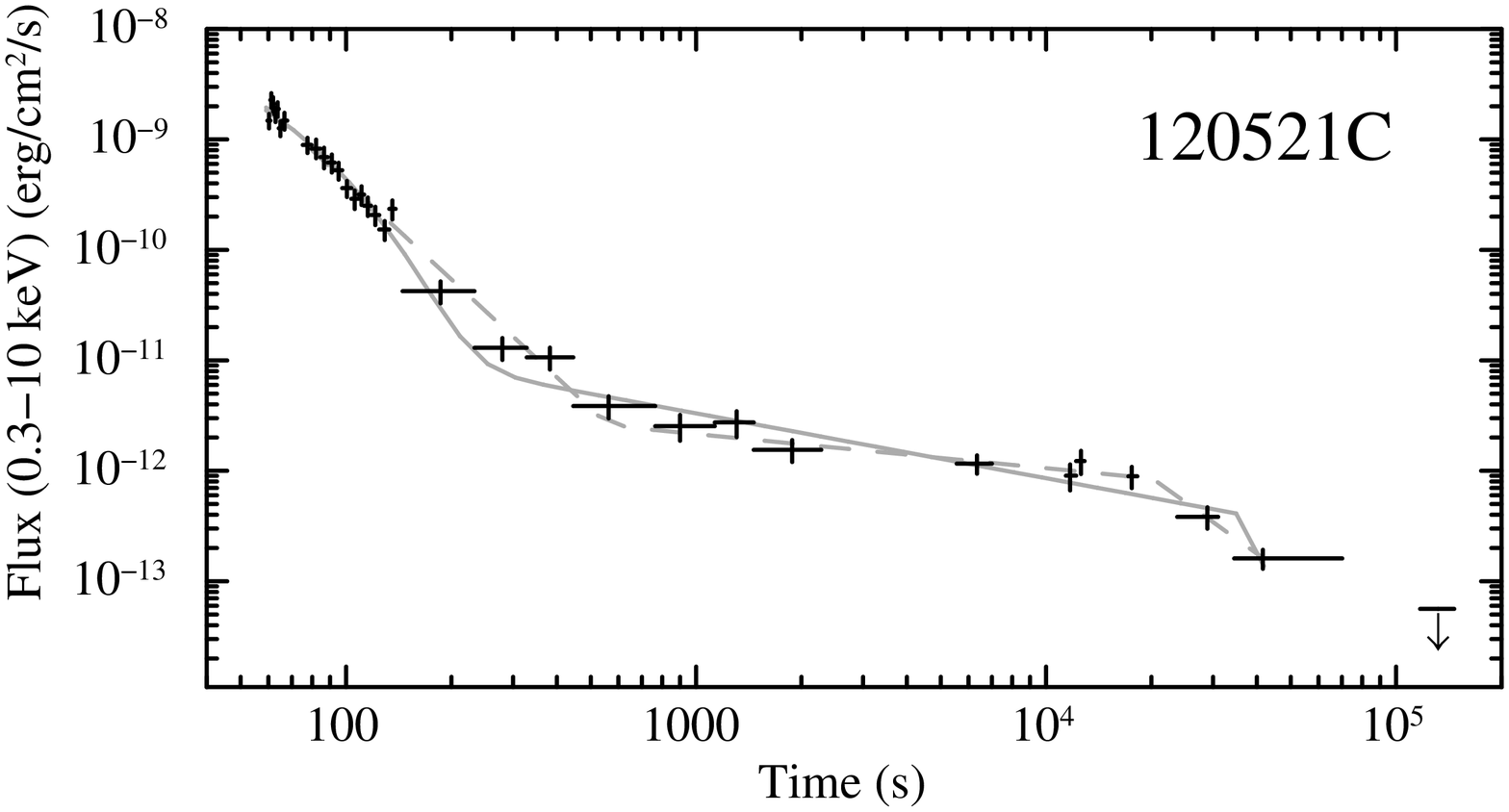}
\includegraphics[width=84mm]{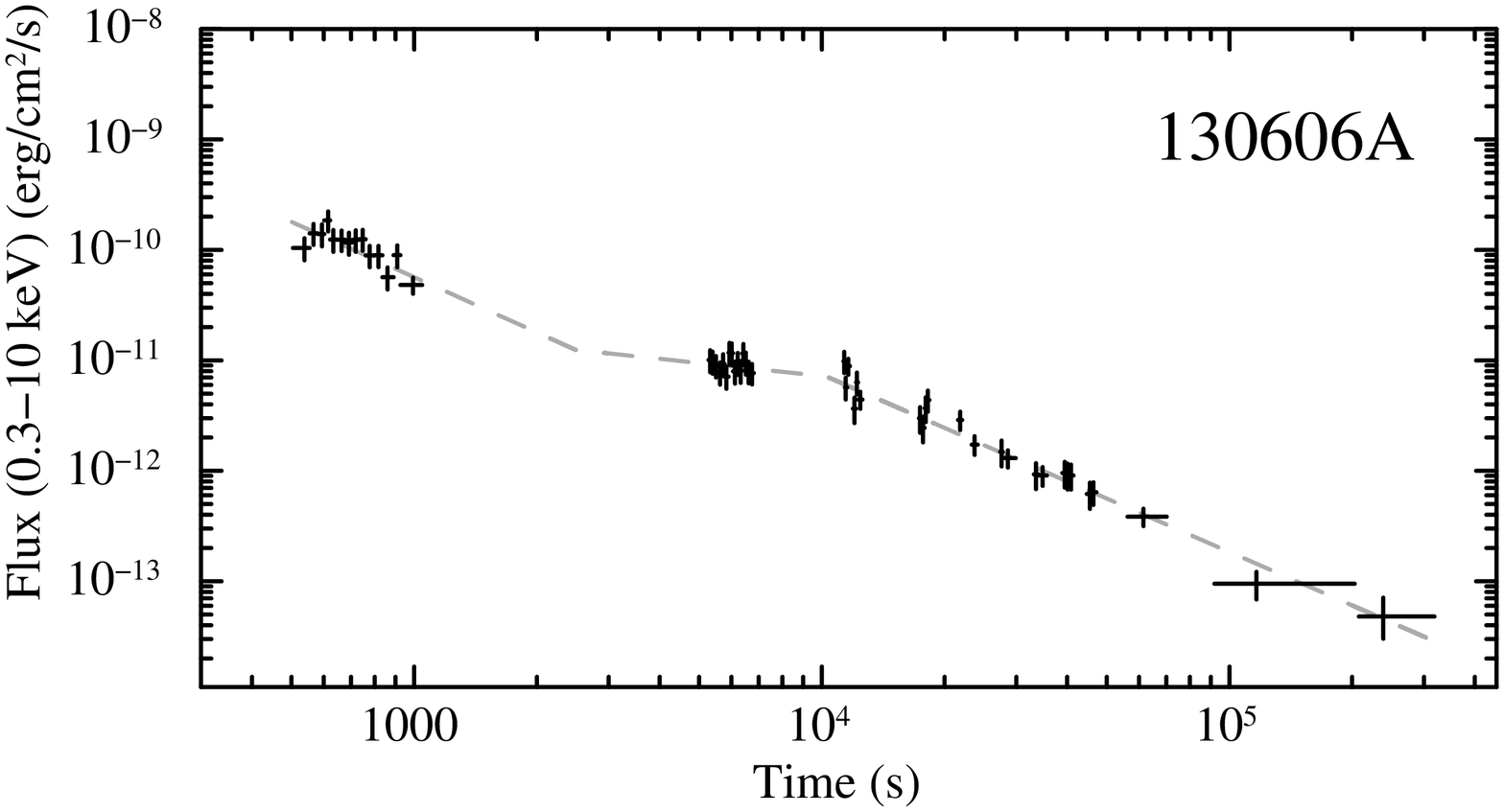}
\caption{
Light curves of following X-ray emissions observed by the XRT (0.3--10\,keV).
Grey dashed and solid lines represent the double-broken PL and the exponential decay plus one-time broken PL function, respectively.
}
\label{fig:2}
\end{figure}

\begin{figure}
\includegraphics[width=84mm]{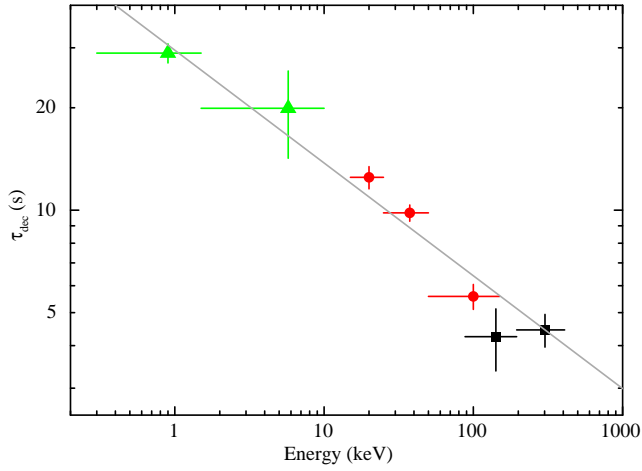}
\caption{
A relation between the energy bandpass and decay time constant $\tau_{\rm dec}$ of \grbB\ in the observer frame (\S\ref{s-anaB}).
Black squares, red circles, and green diamonds are measured values in the WAM, the BAT, and the XRT bandpass, respectively.
}
\label{fig:4.5}
\end{figure}

We tested whether a light curve of a following X-ray emission from an earlier observation, obtained from the \swift\ archive\footnote[3]{http://www.swift.ac.uk} and shown in Figure~\ref{fig:2}, also exhibited an exponential decay.
Afterward, we attempted to identify whether a double-broken PL function better fit ($\chi^{2}/{\rm d.o.f} = 19.7/27$) compared with the combination of an exponential decay and a one-time broken PL ($\chi^{2}/{\rm d.o.f} = 33.0/27$).
We calculated the first and second broken times of the former function to be \TgrbB~$+ (563\pm1)$\,s and \TgrbB~$+ (20167\pm7535)$\,s in the observer frame, respectively
The decay slope indices were $-2.95 \pm 0.09$, $-0.30 \pm 0.07$, and $-2.38 \pm 2.01$ in order of time.
With the XRT, we also divided the light curve into two energy bandpasses of 0.3--1.5\,keV and 1.5--10\,keV, and fit them with the latter function to identify the decay time constant $\tau_{\rm dec}$ in the soft X-ray band.
Although the latter function did not provide the best model fit, the energy dependency of the decay time constant revealed good linearity from an energy band of 0.3--409\,keV in Figure~\ref{fig:4.5}, and a decay index of $\delta_{\tau_{\rm dec}} = -0.33\pm0.03$ was obtained.

\subsubsection{Spectral Analyses of \grbB}

\begin{figure}
\includegraphics[width=84mm]{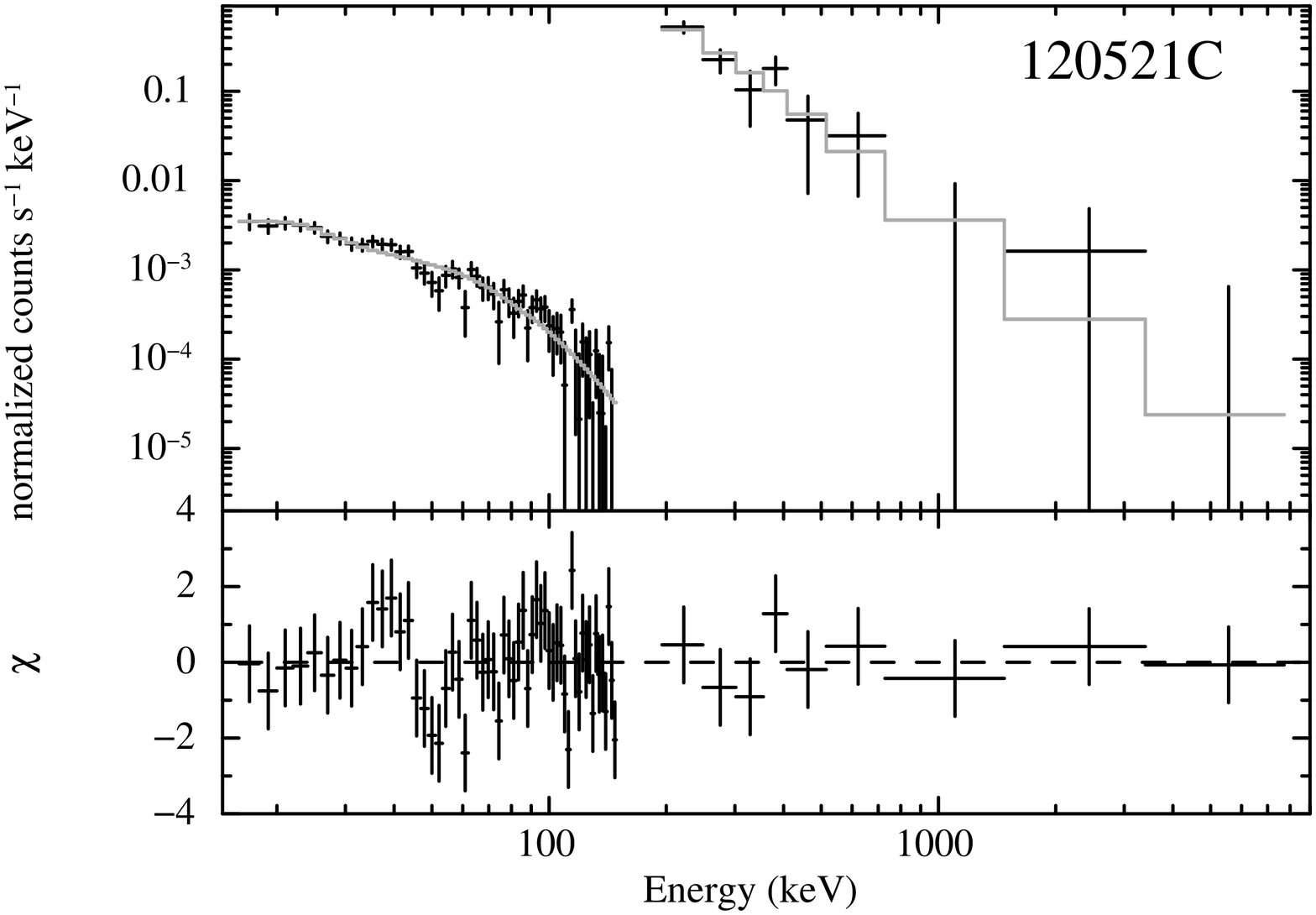}
\includegraphics[width=84mm]{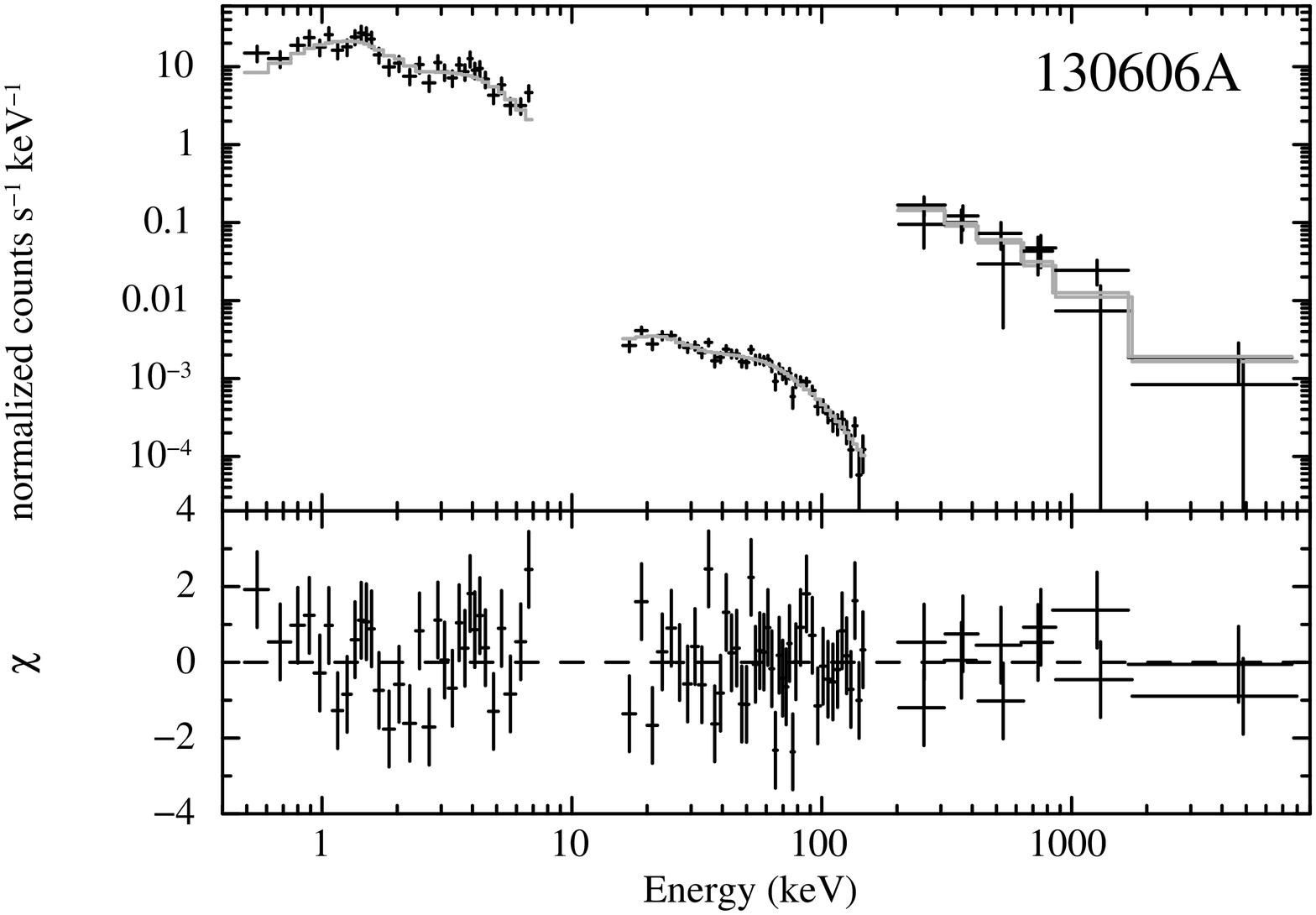}
\caption{
(Left panel) BAT and WAM spectra of \grbB, and (right panel) XRT, BAT, and WAM spectra of region\,C of \grbC.
Gray lines are the best fit Band models.
}
\label{fig:3}
\end{figure}

Time-averaged spectra of WAM-1 and the BAT were accumulated from \TgrbB\ $-2.111$\,s to \TgrbB\ $+10.889$\,s and \TgrbB\ $-2.500$\,s to \TgrbB\ $+11.336$\,s, respectively, by covering the entirety of the WAM $T_{90}$.
A background spectrum of WAM-1 was estimated from an average between (\TgrbB\ $- 145.111$\,s to \TgrbB\ $- 5.111$\,s) and (\TgrbB\ $+ 14.889$\,s to \TgrbB\ $+ 154.889$\,s).
First, the time-averaged spectra were fit with a simple model of the PL function; a photon index was the obtained, which was $\alpha = -1.71\pm0.10$ ($\chi^{2}/{\rm d.o.f}=83.2/64$).
Second, we applied more complex models of a PLE and Band model \citep{band1993}, and these models revealed better test statistics compared with that of the PL.
The parameters obtained from the PLE model were $\alpha = -1.40^{+0.21}_{-0.29}$ and \epeak$ = 115^{+129}_{-37}$\,keV with $\chi^{2}/{\rm d.o.f}=70.1/63$, and those obtained from the Band model were  $\alpha = -1.31^{+0.39}_{-0.28}$, $\beta < -2.20$, and \epeak $= 97^{+121}_{-30}$\,keV with $\chi^{2}/{\rm d.o.f}=69.5/62$.
Figure~\ref{fig:3} shows the fitting results with the Band model.

We further extracted the 1-s peak spectra.
The extracted time region ranged from \TgrbB\ $-1.110$\,s to \TgrbB\ $- 0.110$\,s, which is the brightest interval in the WAM light curve (Figure~\ref{fig:1}).
The peak spectra were fit with the PLE of $\alpha = 0.98^{+0.17}_{-0.22}$ and \epeak$ = 579^{+1656}_{-334}$\,keV with $\chi^{2}/{\rm d.o.f}=46.7/50$, and the Band model of $\alpha = -0.94^{+0.53}_{-0.17}$, $\beta < -1.57$, and \epeak $= 544^{+1620}_{-410}$\,keV with $\chi^{2}/{\rm d.o.f}=46.6/48$.
Its peak flux was $1.41^{+2.34}_{-0.83}\times10^{-6}$\,\ergscm\ in the 1\,keV--10\,MeV band.
The derived parameters of the Band model are shown in Table~\ref{tab:3}.

We also checked XRT spectra of following X-ray emissions, and the data were provided from the UK \swift\ science data centre \citep{evans2009}.
The spectra were extracted from the three intervals from \TgrbB\ $+145$\,s to \TgrbB\ $+559$\,s, from \TgrbB\ $+559$\,s to \TgrbB\ $+18221$\,s, and from \TgrbB\ $+23685$\,s to \TgrbB\ $+70175$\,s which are defined from the fitting results of the light curve with double-broken function (\S\ref{s-anaB-l}), and they were fitted with PL models of photon indexes of $-1.64 \pm0.27$ ($\chi^{2}/$d.o.f = 55.8/72), $-1.71^{+0.21}_{-0.20}$ (91.2/107), and $-1.87^{+0.37}_{-0.35}$ (39.6/47), respectively.
Then we employed fixed Galactic absorption column density of $N^{\rm Gal}_{\rm H} = 1.01 \times 10^{20}$\,cm$^{-2}$ \citep{kalberla2005} and added an extra-galactic absorption with a thawed column density parameter.

%%%  GRB 130606A  %%%
\subsection{GRB\grbC}\label{s-anaC}

%- GRB 130606A -%
The BAT was also triggered by \grbC\ \citep{ukwta2013} on June 6, 2013, at 21:04:39.020 (UT) (\TgrbC, hereafter) in the location of (RA, Dec)(J2000) = (16h37m37s, + 29 d 47' 27"), and goaded many astronomical instruments to observe the following afterglow emissions.
\grbC\ was investigated with optical observatory deeply \citep{castro-tirado2013arXiv,chornock2013,totani2014,hartoog2015}, and its redshift was determined at $z \sim 5.91$.
In particular, \citet{totani2014,totani2016} suggested that the re-ionization state has not completed at the redshift, measuring the IGM neutral fraction (Table~\ref{tab:0}).
Therefore this GRB is good opportunity to get a hint of first generation stars, although the reported redshift is not farther than $z = 6$.

\subsubsection{Light Curves of \grbC}\label{s-anaC-l}

Figure~\ref{fig:1} shows the temporal variations of a part of the prompt emission observed by the three instruments.
In the BAT light curve (15--150\,keV), three intervals were considerably brighter than in the 5.0\,$\sigma$ level from background variations.
We defined the three intervals as region~A (\TgrbC$ - 1.424$\,s to \TgrbC$ +1.904$\,s), region~B (\TgrbC$ + 102.000$\,s to \TgrbC$ +104.304$\,s), and region~C (\TgrbC$ + 152.688$\,s to \TgrbC$ +163.696$\,s).
In the WAM light curve, only region~C was detected successfully with the same criteria, and its $T_{90}$ duration was $9 \pm 2$\,s.
However the $T_{90}$ duration was obviously shorter compared with the light curves of \swift.
We therefore tried to estimate the intrinsic duration of the emission activity using X-ray light curve.

An X-ray light curve of the emissions that followed, published in the \swift\ UK archives\footnote[3]{http://www.swift.ac.uk}, displays a similar decay to that of \grbB, which also fit well with the double-broken PL functions, with $\chi^{2}/{\rm d.o.f} = 41.0/48$ (Figure \ref{fig:2}).
The broken times were thus \TgrbC~$+(2491\pm1694)$\,s and \TgrbC~$+(10244\pm3694)$\,s, and the decay slope indices were $-1.66 \pm 0.42$, $-0.39 \pm 0.80$, and $-1.61 \pm 0.08$ in order of time.

\subsubsection{Spectral Analyses of \grbC} \label{s-anaC-s}

We extracted the spectra from the regions~A and B and performed spectral fitting with only the BAT data sets because the two intervals were detected substantially though only the BAT, but not covered by the XRT.
The PL models of photon indexes of $-1.58 \pm 0.21$ and $-1.29 \pm 0.38$ represented the spectra well, respectively, and the obtained parameters are listed in Table~\ref{tab:3}.
We examined the PLE model instead, and found that the peak energies were not determined because of the low statistics in the high-energy range.

Region~C was evidently detected by all three instruments (Figure~\ref{fig:1}); we therefore extracted the energy spectra from all the instrument data, and performed simultaneous spectral fitting.
To prepare a broadband spectral fitting of region~C simultaneous with the XRT, we derived an extra galactic absorption $N^{\rm ext}_{\rm H}$ of $\sim4.89\times10^{22}$\,cm$^{-2}$ from an averaged spectrum of the XRT, which was accumulated from the entirety of the WT data set (\TgrbC\ $+78.707$\,s -- \TgrbC\ $+ 502.207$\,s) and fit with the PL model (Table~\ref{tab:3}).
We then assumed a Galactic absorption of $N^{\rm Gal}_{\rm H} = 2.00 \times 10^{20}$\,cm$^{-2}$ \citep{kalberla2005} as a fixed parameter.
For the WAM data, we used WAM-1 and WAM-2, which were exposed to the GRBs compared with the other sides.
The extracted intervals of the BAT and the XRT were identical to those defined in region~C (\TgrbC$ + 152.688$\,s to \TgrbC$ +163.696$\,s), whereas those of the WAM were between \TgrbC $+ 152.330$\,s and \TgrbC $+ 163.330$\,s.
Although these spectra were accumulated from slightly different intervals, the differences are only 3\% of the total extracted exposure time.
Therefore, we regarded the differences to be negligible below spectral fitting.
The background spectra of the WAM were estimated to have accumulated spectra from \TgrbC\ $- 57.671$\,s to \TgrbC\ $+ 142.329$\,s, and from \TgrbC\ $+ 174.329$\.s to \TgrbC\ $+ 374.329$\,s.
We employed the XRT background spectrum produced for average spectrum fitting as the background spectrum of region~C and the1-s peak interval, as shown in the following paragraph.
The galactic and extra galactic absorption were fixed at those values.
The spectra better fit the Band model with $\alpha$ of $-0.78 \pm 0.12$, $\beta$ of $-1.56^{+0.22}_{-0.29}$, and $E_{\rm peak}$ of $175^{+80}_{-44}$\,keV ($\chi^{2}/{\rm d.o.f} = 101.5/81$) compared with the PL (134.7/83) and the PLE models (120.3/82).
Figure~\ref{fig:3} shows the fitting results with the Band model, and the parameters obtained from the Band model are listed in Table~\ref{tab:3}.

The 1-s peak spectra were extracted from (\TgrbC\ $+160.374$\,s to \TgrbC\ $+161.374$\,s), which is a part of region~C.
The spectra were represented by the PLE model of $\alpha$ of $-0.87^{+0.26}_{-0.18}$ and \epeak\ of $369^{+483}_{-184}$\,keV ($\chi^{2}/{\rm d.o.f}=60.2/54$) and the Band model with $\alpha$ of $-0.72^{+0.148}_{-0.33}$, $\beta$ of $< -1.78$, and $E_{\rm peak}$ of $221^{+566}_{-104}$\,keV ($\chi^{2}/{\rm d.o.f} = 59.4/53$).
The Band model revealed a flux of $8.95^{+12.03}_{-4.40} \times10^{-7}$\,\ergscm\ (1\,keV--10\,MeV).
The other obtained parameters are listed in Table~\ref{tab:3}.

We also performed spectral analysis of XRT spectra of following parts of from \TgrbC\ $+599$\,s to \TgrbC\ $+1000$\,s, from \TgrbC\ $+5271$\,s to \TgrbC\ $+6813$\,s, and from \TgrbC\ $+21598$\,s to \TgrbC\ $+$29891\,s.
We defined the three intervals from the results of the light curve fitting with the double-broken PL function (\S\ref{s-anaC-l}).
The XRT spectra were represented by PL models of photon indexes of $-1.51 \pm0.15$ ($\chi^{2}/{\rm d.o.f} =154.9/172$), $-1.69^{+0.14}_{-0.13}$ ($147.9/186$), and $-1.81 \pm0.26$ ($72.5/82$), respectively, in order of observation times, and then we employed fixed Galactic absorption column density of $N^{\rm Gal}_{\rm H} = 2.0 \times 10^{20}$\,cm$^{-2}$ \citep{kalberla2005} and thawed extra-galactic absorption column density as fitting parameters.

%%%   Discussion   %%%
\section{Discussion} \label{s-dis}

We presented our temporal and spectral analyses of the prompt emissions of two GRBs, (i.e., \grbB\ and \grbC), at high-redshift $z\sim6$ and $z\sim5.91$, respectively.
The results revealed that we succeeded in determining these spectral shapes as listed in Table~\ref{tab:3}, because of the broadband simultaneous spectra with an energy band ranging from at least 15\,keV--7.7\,MeV.

\subsection{Temporal Characteristics} \label{s-disT}

\begin{figure}
\includegraphics[width=84mm]{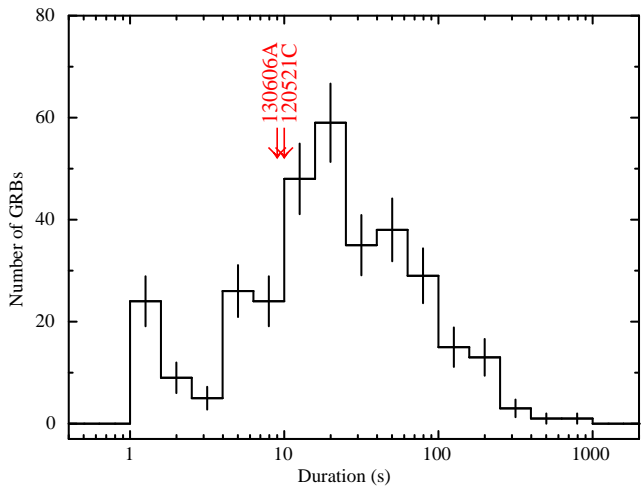}
\caption{
$T_{90}$ distribution constructed by the WAM with observed GRBs with a 1-s time resolution data presented in \citet{ohmori2016}.
The $T_{90}$ values are in the observer frame.
Our results were calculated using only WAM light curves of Figure~\ref{fig:1}.
}
\label{fig:3.5}
\end{figure}

The $T_{90}$ durations derived from the WAM light curves (\S\ref{s-anaB-l} and \S\ref{s-anaC-l}) were $10 \pm 4$\,s (\grbB) and $9 \pm 2$\,s (\grbC).
Figure~\ref{fig:3.5} shows the $T_{90}$ distribution, which was constructed by the WAM with observed GRBs with a 1-s time resolution data format \citep{ohmori2016}, and represented with a single lognormal Gaussian centered at $24.4^{+2.9}_{-2.6}$\,s, excluding the 1-s bin.
Compared with the $T_{90}$ distribution, our results were shorter than the peak of the distribution.
However, the durations of the prompt emissions obviously differed in each energy band (Figure~\ref{fig:1}), and thus, we attempted to evaluate the durations though a different approach involving $T_{90}$.

% Duration of 120521C %
Both XRT light curves (Figure~\ref{fig:2}) assume similar shapes, as represented by the double broken PL function (\S \ref{s-anaB}).
Specifically, the early part of the light curve of \grbB\ also exhibited exponential decay, and these decay constants were correlated with those of a higher bandpass of the BAT and the WAM (Figure~\ref{fig:4.5}).
The derived energy dependency index of $\delta_{\tau _{\rm dec}} = -0.33 \pm 0.03$ is consistent with the values reported in several previous studies of FRED GRBs of $-0.33\pm0.23$ \citep{peng2012}, $-0.34\pm0.12$ \citep{tashiro2012}, and $-0.29\pm0.12$ \citep{tashiro2014}, although these dispersions are large.
We therefore considered the emission until the first break time of \TgrbB~$+563\pm1$\,s at the longest duration not to be afterglow dominant, but as the tail of the prompt emission.
Furthermore, we assumed duration of X-ray radiation in the rest frame using energy dependency index of $\delta_{\tau_{\rm dec}} \sim -0.3$ and the equation of
\begin{equation}
t_{\rm rest} = \cfrac{{t}_{\rm obs}}{(1+z)} \times \left(\cfrac{E_{\rm obs}(1+z)}{E_{\rm rest}}\right)^{-\delta_{\tau_{\rm dec}}},
\end{equation}
where $t_{\rm rest}$ and $E_{\rm rest}$ are duration and energy of photons in the rest frame of the GRB, respectively, and $t_{\rm obs}$ and $E_{\rm obs}$ are those of in the observer frame.
Then the duration of $t_{\rm obs} \sim 563$\,s corresponds to $t_{\rm rest}\sim144$\,s.
The value of the energy dependency index is insufficient to be explained with synchrotron and/or inverse-Compton cooling predicting one of $\delta_{\tau_{dec}} = -1/2$, thus implying the presence of an additional component and/or a geometrical effect in the emission region \citep{tashiro2014}.
However, the prompt emissions of \grbB\ were too faint to investigate the spectral evolution.
We further calculated the energy dependency index of the pulse width $w$ of $-0.47 ^{+0.15}_{-0.16}$ excluding the XRT light curve because it does not include the pulse rise interval that enables a comparision against other GRBs \citep[e.g.][]{peng2012,shenoy2013}.
Although the dispersions of the index value are large, it is comparable with the reported distribution of $-0.32 \pm 0.17$ \citep{peng2012}.

\citet{hartoog2015} reported that the optical to X-ray energy spectrum of the \grbC\ afterglow are well reproduced with a single PL model of a photon index of $-2.02$.
The photon index is consistent with the value of $-1.81 \pm 0.26$ that we observed after second break time (\S\ref{s-anaC-s}; \TgrbC$>10244$\,s).
Moreover the X-ray light curve in its interval showed a simple PL decay.
We therefore conclude that the dominant component is not prompt emission but afterglow after the second break.
Also the spectral shapes of three intervals after \TgrbC\ $+599$\,s were not different each other, but the first interval shows the most similar photon index with the average XRT spectrum of the prompt emission.
We finally defined the interval until the first break time of the fitted double-broken PL function dominates the prompt emission for \grbC\ like that of \grbB.

% Duration of 130606A %
Although theoretical studies \citep[e.g.][]{suwa2011} have predicted that the prompt emissions have a long duration of approximately $1000$\,s in the rest frame of GRBs produced from Pop. \three\ stars, an approximately 80\,s (or $\sim 144$\,s) duration of \grbB\ in the rest frame is insufficient for such a prediction.
The duration is comparable with the expected duration of $49$\,s from a progenitor of the Wolf-Rayet star \citep{suwa2011}.
Although the duration of \grbC, approximately $360$\,s in the rest frame, is substantially longer than that of \grbB\ significantly, it is still insufficient to compare with the predicted value for the Pop. \three\ stars.
The $t_{\rm burst}$ estimation of 72\,s in the rest frame for \grbC\ \citep{zhang2014} also supports this conclusion.

\subsection{Radiated Energy Correlations} \label{s-disE}

\begin{figure}
\includegraphics[width=84mm]{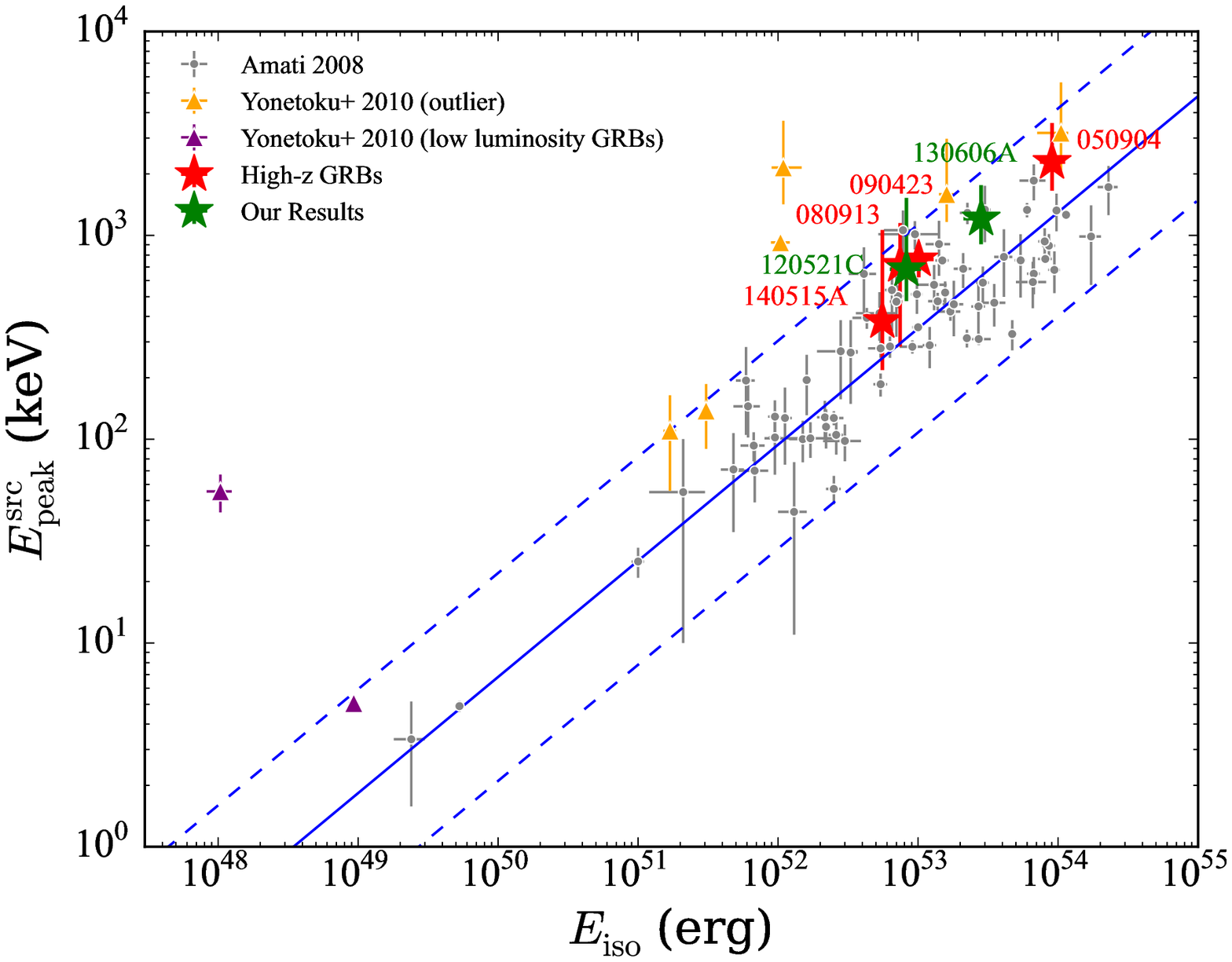}
\includegraphics[width=84mm]{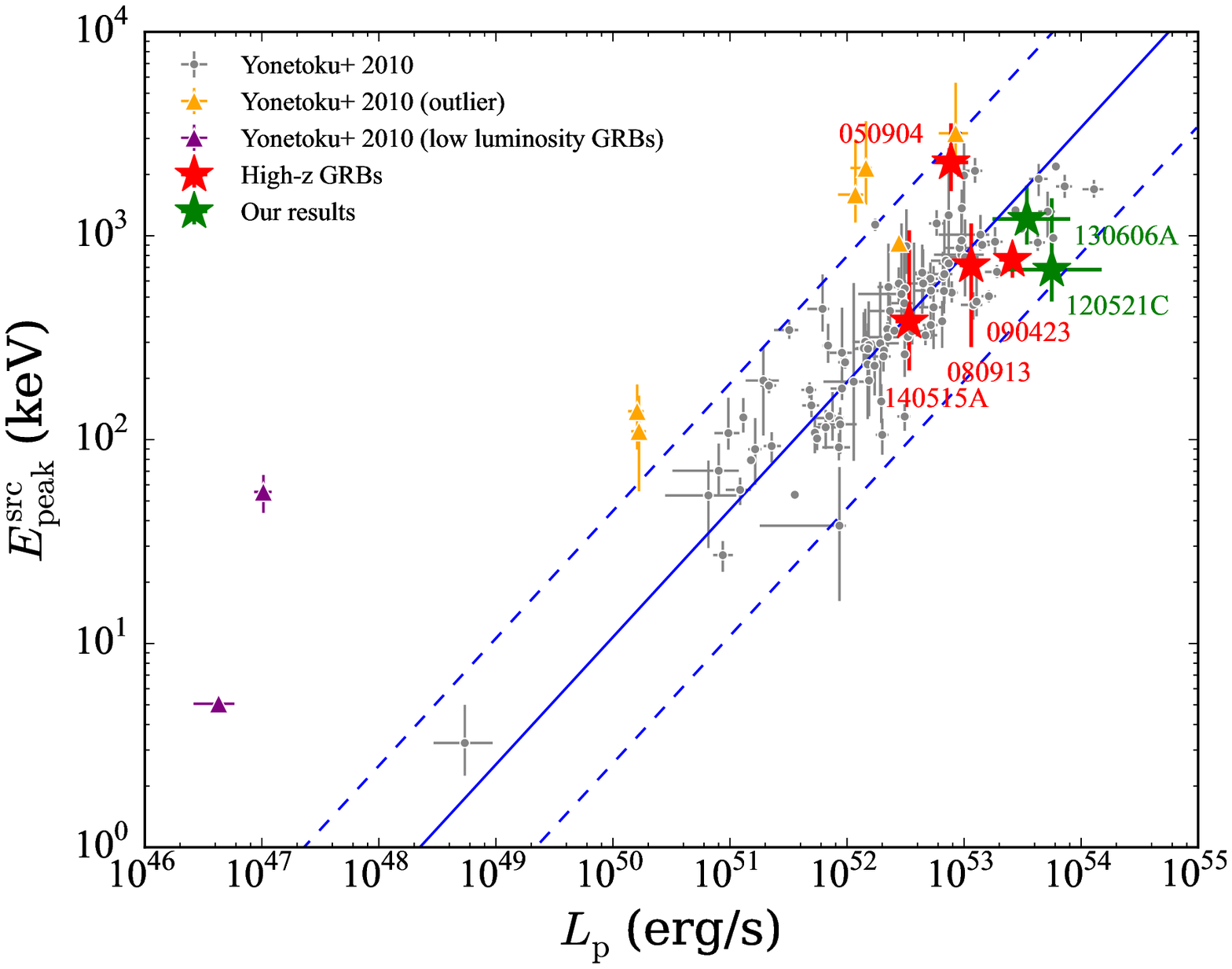}
\includegraphics[width=84mm]{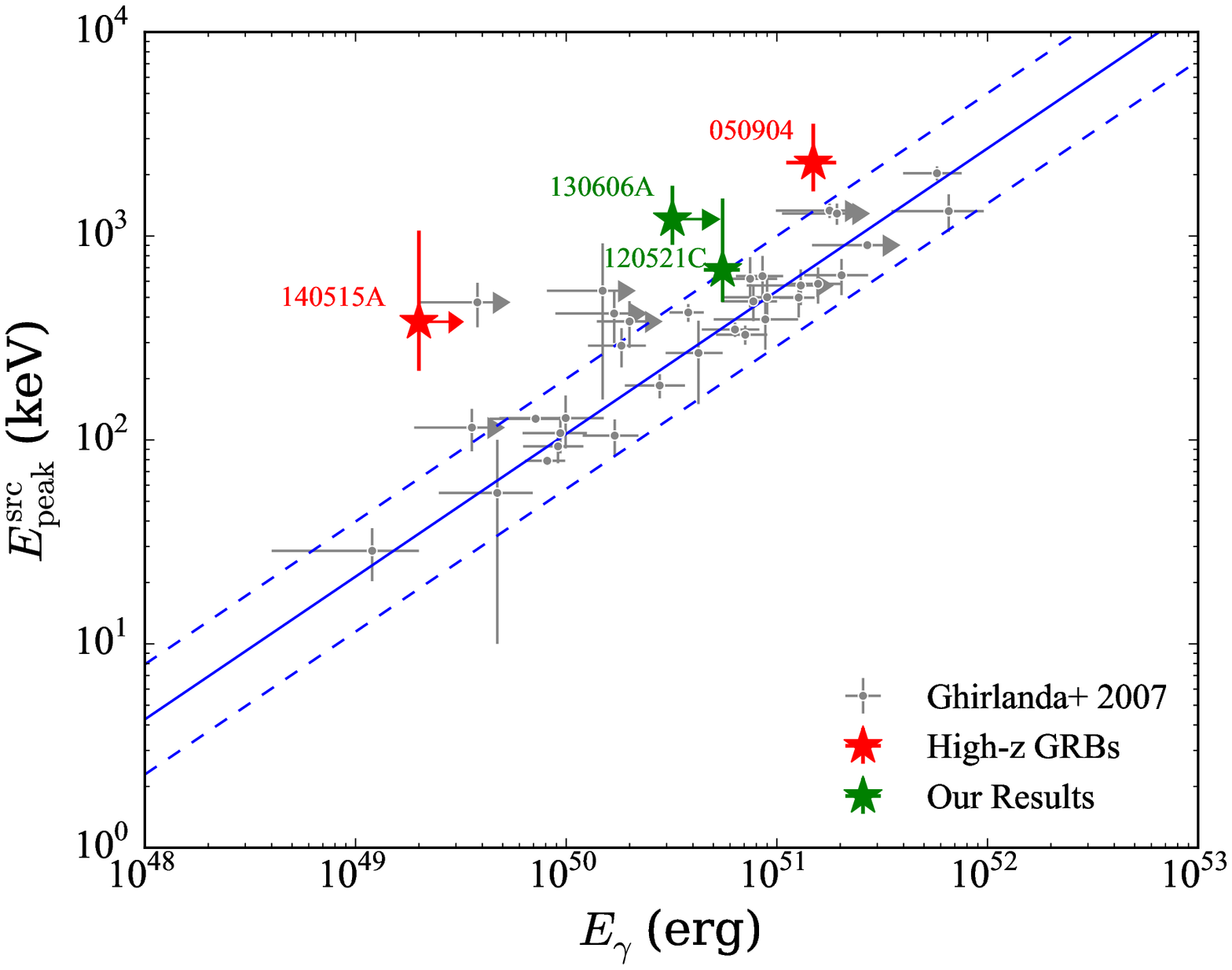}
\caption{
The three notable empirical relations.
Gray points indicate the data of \citet{amati2008}, \citet{yonetoku2010}, and \citet{ghirlanda2007}, and blue solid and dashed lines are fitted power-law functions and those $3\sigma$\ regions derived in each paper, respectively.
Green stars indicate our results.
Red stars are the other high-redshift GRBs, and these values are refereed from several previous studies of \citet{ghirlanda2007,sugita2009,yonetoku2010,melandri2015}.
Yellow and purple diamonds are defined as outliers and low luminosity GRBs in \citet{yonetoku2010}.
}
\label{fig:5}
\end{figure}

Regarding the emission mechanism compared with typical GRBs, we tested our results by using two well-known relations, namely the Amati \citep[\epeakrest-\eiso;][]{amati2002} and Yonetoku \citep[\epeakrest-\lpeak;][]{yonetoku2004} relations.
We calculated the intrinsic spectral energy peak \epeakrest, the isotropic radiated energy \eiso, and the 1-s peak luminosity \lpeak.
The energy bands of \eiso\ and \lpeak\ are defined as the range of 1\,keV--10\,MeV.
Afterward, the luminosity distances of 57.7\,Gpc (\grbB) and 56.7\,Gpc (\grbC) were calculated from each redshift of $z\sim6$ and $5.91$, respectively.
The parameters derived from our samples are as follows: \epeakrest $= 682^{+845}_{-207}$\,keV, \eiso $= (8.25 ^{+2.24}_{-1.96})\times 10^{52}$\,\erg, and \lpeak $= (5.62^{+9.33}_{-3.31}) \times 10^{53}$\,\ergs\ for \grbB; and \epeakrest\ $= 1209^{+553}_{-304}$\,keV, \eiso $=(2.82^{+0.17}_{-0.71})\times 10^{53}$\,\erg, and \lpeak $= (3.44^{+4.63}_{-1.69}) \times 10^{53}$\,\ergs\ for \grbC.
As shown in Figure~\ref{fig:5}, according to the Amati relation, both of our samples were consistent within the 2.0\,$\sigma$ (\grbB) and 1.7\,$\sigma$ (\grbC) levels of the correlation PL function reported by \citet{amati2008}.
Moreover, the \epeakrest-\lpeak\ value of \grbC\ fits the Yonetoku relation \citep{yonetoku2010} within the 0.8\,$\sigma$ level, but the mean value of \grbB\ deviated by 2.6\,$\sigma$.
The result of \grbB\ possibly supports the prediction of the fireball dissipation in photosphere of large radius Pop. \three\ star.

Certain objects have already been reported as outliers in the Amati and/or Yonetoku relation \citep{yonetoku2010}, and these points show an inclination of higher \epeakrest\ values compared with the mean of the relation, as shown in Figure~\ref{fig:5}.
Because the point of \grbB\ is in the lower \epeakrest\ and higher \lpeak\ side, it might imply a new type of outlier in the Yonetoku relation.
Although \grbA, the redshift of which is $z=6.295\pm0.002$ \citep{kawai2006}, was identified as an outlier in \citet{yonetoku2010}, a result of a detailed and broadband spectral analysis \citep{sugita2009} indicated an inlier within the 3\,$\sigma$ level.
Thus, we cannot reject the statistical effect caused by a large uncertainty and declare that \grbB\ is a new outlier.

We further calculated the geometrically corrected radiated energy \egamma\ of \grbB\ and \grbC, and compared other GRBs against the Ghirlanda relation \citep[\epeakrest-\egamma;][]{ghirlanda2004,ghirlanda2007}.
Afterward, we assumed an energy conversion efficiency of 0.2 to radiate, with an ambient density of 3\,cm$^{3}$, respectively.
According to the jet break time of \grbB\ at 7\,d \citep{laskar2014}, as revealed by radio observations, the jet opening angle and \egamma\ were estimated to be approximately $6.6^{\circ}$ and $(5.5 ^{+1.1}_{-1.0}) \times 10^{50}$\,\erg, respectively.
To convert the measured jet break time to the jet opening angle, we then used the formula of \citet{sari1999,frail2001}.
For \grbC, a jet break time has not been determined through any follow-up observations, and an obvious break has not been seen in the reported optical light curve of the $z'$-band \citep{trotter2013} until 1.3\,d.
Therefore, we assumed the jet break time to be $>1.3$\,d, and estimated \grbC\ to have a jet opening angle of $>3.2^{\circ}$ and an \egamma\ value of $>3.2 \times 10^{50}$\,\erg.
As shown in Figure~\ref{fig:5}, the estimated \egamma\ values were plotted according to the Ghirlanda relation, and both points in the figure fit the relation.
Furthermore, assuming a jet opening angle of $7^{\circ}$, which is a median of a distribution \citep{fong2012} that indicates a jet break time of 12.3\,d, \egamma\ is estimated as $2.26\times10^{51}$\,\erg, and the value is an inlier of the correlation within the 1.2\,$\sigma$ level.
Although \grbA\ is outside the 3\,$\sigma$ level, after applying dusty circumburst density around the GRB, it also fits the Ghirlanda relation \citep{sugita2009}.

In conclusion, our samples fit the three notable empirical relations.
Because \grbC\ has been implied as having radiated from the reionization epoch \citep{totani2014}, our findings revealed that the relations conformed not only to low-redshift GRBs, but also to high-redshift GRBs from the epoch.
According to our results of insufficient durations and the fitting with the three relations, these properties are dissimilar to those of the prediction of Pop. \three\ stars, but apply instead to ordinary stars such as Walf-Rayet stars.

%%----  Acknowledgments  ----%%
\section*{acknowledgments}
We are grateful to all the \suzaku-WAM team members.
We also thank Norisuke Ohmori for providing useful information of WAM $T_{90}$ distribution.
This work is partly supported by the Ministry of Science and Technology of Taiwan grants MOST 104-2112-M-008-011- and 105-2112-M-008-013-MY3 (YU).
This work made use of data supplied by the UK Swift Science Data Centre at the University of Leicester.

%%%%  Table-1  %%%%
\begin{table}
\caption{High-redshift GRBs}
\label{tab:0}
\scriptsize
%\footnotesize
\begin{tabular}{@{}lcccccccc}
\hline
& Redshift $z$ & \shortstack{Prompt\\observation} & $T_{90}$ (BAT) & $t_{\rm burst}$ & Model & \epeakrest & \eiso & \shortstack{the IGM\\neutral fraction}\\
&                      &                                   & (s) & (s) & & (keV) & (\erg) & $n_{H_{I}}/n_{H}$ \\
\hline
050904   & $6.295$ $^a$    & BAT, XRT, WAM & $225 \pm 10 $ $^b$ &  $\geqslant3.15\times10^{5}$ $^c$ & Band & $2291^{+1263}_{-634}$ $^d$ & $(1.04 ^{+0.25}_{-0.17})\times10^{54}$ $^d$ & $<0.6$$^e$\\
080913   & $6.695$ $^f$      & BAT, XRT &  $8 \pm 1$ $^g$ & --- & Band (fixed $\beta$) & $716.4\pm431.7$ $^h$ & $(7.44\pm0.80)\times10^{52}$ $^h$ &  --\\
090423   & $8.1$ $^i$ & GBM$^j$, BAT, XRT &  $10.3 \pm 1.1$ $^k$ &  ---  & Band (fixed $\beta$) & $762.6\pm139.5$ $^h$ & $(1.01\pm0.28)\times10^{53}$ $^h$ & --\\
090429B & $\sim9.4$ $^l$ & BAT, XRT, &  $5.5 \pm 1.0$ $^m$ & --- & -- & -- & -- & --\\
120521C & $\sim6$ $^n$   & BAT, XRT, WAM &  $26.7 \pm 4.4$ $^o$ & --- & Band & $682^{+845}_{-207}$ $^p$ & $(8.25 ^{+2.24}_{-1.96})\times 10^{52}$ $^p$ & --\\
130606A & $5.913$ $^q$ & BAT, XRT, WAM, Konus &  $276.58 \pm 19.31$ $^r$ & $(4.98\pm0.09)\times10^{2}$ $^c$ & Band & $1209^{+553}_{-304}$ $^p$ & $(2.82^{+0.17}_{-0.71})\times 10^{53}$ $^p$ & 0.1--0.5 $^s$\\
140515A & $6.32$ $^t$      & BAT, XRT & $23.4 \pm 2.1$ $^u$  & --- & PLE  & $379.7 ^{+681.7}_{-161.3}$ $^v$ & $(5.8 \pm0.6) \times 10^{52}$ $^v$ & $<0.002$ $^v$\\
\hline
\multicolumn{9}{l}{
\shortstack{
References; 
$^a$\cite{kawai2006}, % a
$^b$\cite{sakamoto2005}, % o
$^c$\cite{zhang2014}, % v 
$^d$\cite{sugita2009},  % b
$^e$\cite{totani2006}, % c
$^f$\cite{greiner2009}, % d
$^g$\cite{stamatikos2008}, % p
$^h$\cite{yonetoku2010},\\ % e
$^i$\cite{salvaterra2009}, % f
$^k$\cite{palmer2009}, % q
$^l$\cite{cucchiara2011}, % g
$^m$\cite{stamatikos2009}, % r
$^n$\cite{laskar2014}, % h
$^o$\cite{markwardt2012}, % s
$^p$this work, % i
$^q$\cite{chornock2013},\\ % j
$^r$\cite{barthelmy2013}, % t
$^s$\cite{totani2014}, % k
$^t$\cite{chornock2014}, % l
$^u$\cite{stamatikos2014}, % u
$^v$\cite{melandri2015} % m
}}\\ 
\multicolumn{9}{l}{$^j$ GBM is gamma-ray burst monitor onboard \fermi\ satellite.}
\end{tabular}

 \medskip
% Comment is here 
\end{table}
\normalsize

\begin{table}
 \caption{Two FRED charactaristic parameters of \grbB\ in the observer frame.}
 \label{tab:1}
 \begin{tabular}{@{}llccccccc}
  \hline
  Instrument & Bandpass & $\chi^{2}$/d.o.f\,$^a$ & $\tau_{1}$\,(s) & $\tau_{2}$\,(s) & $w$\,(s) & $k$ & $\tau_{\rm rise}$\,(s) & $\tau_{\rm dec}$\,(s)\\
  \hline
  WAM & 88\,keV--7.7\,MeV & 44.4/56 & $0.00 \pm 0.01$ & $3.76 \pm0.73$ & $3.81 \pm2.28$ & $0.99 \pm0.56$ & $0.02 \pm1.55$ & $3.79 \pm1.55$\\
   & 195--409\,keV & 55.3/56 & $0.00 \pm 0.01$ & $4.42 \pm0.93$ & $4.48 \pm0.97$ & $0.98 \pm0.04$ & $0.03 \pm0.49$ & $4.45 \pm0.49$\\
   & 88--195\,keV & 50.4/56 & $0.57 \pm 0.97$ & $3.22 \pm1.26$ & $5.28 \pm2.52$ & $0.61 \pm0.17$ & $1.03 \pm0.88$ & $4.25 \pm0.88$\\
  BAT & 15--150\,keV    & 126.2/113 & $0.17 \pm0.07$ & $8.20 \pm0.58$ & $10.30 \pm0.82$ & $0.82 \pm0.03$ & $1.05 \pm0.36$ & $9.25 \pm0.36$ \\
   & 50--150\,keV            & 122.8/113 & $ 0.30\pm0.18$ & $ 4.62\pm0.78$ & $ 6.56\pm1.22$ & $ 0.70\pm0.06$ & $ 0.97\pm0.47$ & $ 5.58\pm0.47$\\
   & 25--50\,keV              & 144.9/113 & $ 0.39\pm0.18$ & $ 8.32\pm0.86$ & $ 11.34\pm1.33$ & $ 0.73\pm0.04$ & $ 1.51\pm0.54$ & $ 9.82\pm0.54$\\
   & 15--25\,keV              & 87.3/113 & $ 1.03\pm0.65$ & $ 9.94\pm1.36$ & $ 15.04\pm2.48$ & $ 0.66\pm0.06$ & $ 2.55\pm0.94$ & $ 12.49\pm0.94$\\
  XRT & 0.3--10\,keV & 33.0/27 & -- & -- & -- & -- & -- & $28.64 \pm1.52$ \\
   & 1.5--10\,keV & 5.4/6 & -- & -- & -- & -- & -- & $19.93 \pm5.72$ \\
   & 0.3--1.5\,keV & 17.0/29 & -- & -- & -- & -- & -- & $28.94 \pm1.86$ \\
  \hline
  \multicolumn{9}{l}{$^a$ d.o.f is degree of freedom.}
 \end{tabular}

 \medskip
% Comment is here 
\end{table}

\begin{table}
 \caption{List of employed data sets for spectral analyses.}
 \label{tab:2}
 \begin{tabular}{@{}lcccc}
  \hline
  & Instrument & ObsID & Start Time (UT) & Stop Time (UT)\\
  \hline
  \grbB & WAM & 707013010 & 2012-05-20 20:54:08 & 2012-05-22 21:39:12\\
  & BAT & \multirow{2}{*}{00522656000} & 2012-05-21 23:18:15 & 2012-05-21 23:38:17\\
  & XRT &  & 2012-05-21 23:23:30 & 2012-05-21 23:24:38\\
  \hline
  \grbC & WAM & 708021020 & 2013-06-05 05:55:19 & 2013-06-07 23:00:11 \\
  & BAT & \multirow{2}{*}{00557589000} & 2013-06-06 20:50:59 & 2013-06-07 04:40:54 \\
  & XRT & & 2013-06-06 21:06:05 & 2013-06-07 09:29:15\\
  \hline
 \end{tabular}
 
 \medskip
% Comment is here 
\end{table}

\begin{table}
 \caption{Fitting Parameters.}
 \label{tab:3}
 \begin{tabular}{@{}lccccccccc}
  \hline
  & Interval & Instruments & $N^{\rm ext}_{\rm H}$ & $\alpha$ & $\beta$ & \epeak & Flux (15\,keV--5\,MeV) & $\chi^{2}/{\rm d.o.f}$ & Model\\
  & & & [$10^{22}$\,cm$^{-2}$] & & & [keV] & [$10^{-7}$\,erg\,s$^{-1}$\,cm$^{-2}$] & & \\
  \hline
  \grbB & Average & BAT+WAM    & -- & $-1.31^{+0.39}_{-0.28}$ & $<-2.20$ & $97^{+121}_{-30}$ & $1.01^{+0.63}_{-0.25}$ & 53.7/44 & Band\\ 
            & Peak 1\,s & BAT+WAM  & -- & $-0.94^{+0.53}_{-0.17}$ & $<-1.57$ & $545^{+1581}_{-410}$ & $12.91^{+15.33}_{-7.32}$ & 45.6/49 & Band \\ 
  \hline
  \grbC & Average & XRT & $4.89^{+1.78}_{-1.68}$ & $-1.53\pm0.04$ & -- & -- & -- & 308.8/313 & PL \\
            & Region\,A & BAT  & -- & $-1.58\pm0.21$ & -- & -- & $3.79^{+2.94}_{-1.74}$ & 58.4/44 & PL \\
            & Region\,B & BAT  & -- & $-1.29\pm0.38$ & -- &-- & $6.02^{+16.65}_{-3.90}$ & 24.3/20 & PL \\
            & Region\,C & XRT+BAT+WAM  & $4.89$\,(fixed) &$-0.78 \pm0.12$ & $-1.56^{+0.22}_{-0.29}$ & $175^{+80}_{-44}$ & $9.00^{+8.75}_{-3.52}$ & 101.5/81 & Band\\ 
            & Peak 1\,s & XRT+BAT+WAM  & $4.89$\,(fixed) & $-0.72^{+0.48}_{-0.33}$ & $<-1.78$ & $221^{+566}_{-104}$ & $8.08^{+9.25}_{-3.74}$ & 59.4/53 & Band\\ 
  \hline
 \end{tabular}
 
 \medskip
% Comment is here 
\end{table}

%%%  Figures  %%%

\bsp

\end{document}